\documentclass[useAMS,usenatbib]{mn2e}
\usepackage{amsmath,amssymb,graphicx}
\usepackage[english]{babel}
\title[The AM CVn star SDSS\,J1552+3201]{The long-period AM CVn star SDSS\,J155252.48$+$320150.9}
\author[G.\,H.\,A. Roelofs et al.]{G.\,H.\,A.~Roelofs,$^{1,2}$\thanks{E-mail: groelofs@cfa.harvard.edu} P.\,J.~Groot,$^1$ D.~Steeghs,$^{2,3}$ T.\,R.~Marsh$^3$ and G.~Nelemans$^1$\\
$^1$Department of Astrophysics/IMAPP, Radboud University, PO Box 9010, 6500 GL Nijmegen, The Netherlands\\
$^2$Harvard--Smithsonian Center for Astrophysics, 60 Garden Street, Cambridge, MA 02138, USA\\
$^3$Department of Physics, University of Warwick, Coventry CV4 7AL, UK\\
}
\newcommand{\obj}{SDSS\,J1552}
\newcommand{\Porb}{$P_\mathrm{orb}=3376.3\pm 0.3$\,s}
\newcommand{\Teff}{$T_\mathrm{eff}=12,000-15,000$\,K}
\begin{document}
\maketitle

\begin{abstract}
The Sloan Digital Sky Survey has been instrumental in obtaining a homogeneous sample of the rare AM CVn stars: mass-transferring binary white dwarfs.
As part of a campaign of spectroscopic follow-up on candidate AM CVn stars from the Sloan Digital Sky Survey, we have obtained time-resolved spectra of the $g=20.2$ candidate SDSS\,J155252.48$+$320150.9 on the Very Large Telescope of the European Southern Observatory. We report an orbital period \Porb, or $56.272\pm0.005$ min, based on an observed `S-wave' in the helium emission lines of the spectra. This confirms the ultracompact nature of the binary. Despite its relative closeness to the orbital period minimum for hydrogen-rich donors, there is no evidence for hydrogen in the spectra. We thus classify SDSS\,J1552 as a new bona fide AM CVn star, with the second-longest orbital period after V396 Hya ($P=65.5$\,min). The continuum of SDSS\,J1552 is compatible with either a blackbody or helium atmosphere of \Teff. If this represents the photosphere of the accreting white dwarf, as is expected, it puts the accretor at the upper end of the temperature range predicted by thermal evolution models. This suggests that \obj\ consists of (or formerly consisted of) relatively high-mass components.
\end{abstract}


\section{Introduction}

The AM CVn stars are double-degenerate, interacting white dwarf binaries with ultrashort orbital periods (below the orbital period minimum for hydrogen-rich donors). Since they form the evolutionary end-product of several stellar evolutionary scenarios (e.g.\ \citealt{nelemans,podsi} for recent studies) they are of interest as calibrators for binary evolution theory. Furthermore, their importance as \emph{LISA} sources has been noted (e.g.\ \citealt{nelemans2004,stroeer2005,stroeer2006}), being currently the only known observable sources of gravitational waves (e.g.\ \citealt{roelofshst}), as well as their potential for producing (subluminous) supernova-Ia-like explosions (`SN\,.Ia'; \citealt{.Ia}). Although about 20 AM CVn systems are known at present, the majority of these have emerged serendipitously from various surveys over the years, which until recently had made it impossible to study their population and derive fundamental numbers such as their space density, crucial for calibrating predictions from binary stellar evolution theory.

The Sloan Digital Sky Survey (SDSS; \citealt{york}) has provided the first homogeneous sample of six systems \citep{roelofs,anderson,groot} for which a population study could be done \citep{roelofspop}. Table \ref{systems} shows the basic data on these six objects. In order to characterise this sample, we have pursued follow-up observations of the new (candidate) systems reported in the aforementioned papers. In this paper we report our results of an extensive set of observations of one of these, SDSS J155252.48+320150.9 (hereafter \obj), that we obtained with the Very Large Telescope of the European Southern Observatory, Chile.

\begin{table*}
\begin{center}
\begin{tabular}{l l l l l l l l}
\hline
SDSS			&$u$	&$g$	&$r$	&$i$	&$z$	&$P_\mathrm{orb}$ (min)	&Reference\\
\hline
\hline
J012940.05$+$384210.4	&19.66&19.81&20.04&20.23&20.60&-			&\citet{anderson}\\
J092638.71$+$362402.4	&18.71&18.96&19.19&19.39&19.39&$28.31\pm 0.01$	&\citet{anderson,marsh0926}\\
J120841.97$+$355025.2	&18.80&18.77&18.94&19.09&19.17&-			&\citet{groot}\\
J124058.03$-$015919.2	&19.46&19.56&19.79&20.02&20.14&$37.36\pm 0.01$	&\citet{roelofs}\\
J141118.31$+$481257.6	&19.28&19.35&19.51&19.72&19.82&$46\pm 2$		&\citet{anderson,groot}\\
J155252.48$+$320150.9	&20.13&20.23&20.32&20.41&20.57&$56.272\pm 0.005$&This paper; \citet{anderson}\\
\hline
\end{tabular}
\caption{Overview of the six AM CVn systems (or candidates) discovered from the SDSS-I, including coordinates, $ugriz$ magnitudes, and current measurements of their orbital periods if available.}
\label{systems}
\end{center}
\end{table*}

\section{Observations and data reduction}

\begin{table}
\begin{center}
\begin{tabular}{l l r}
\hline
Date		&UT		        &Number of\\
                &                       &exposures\\
\hline
\hline
2006/05/01	&06:41--07:30	        &13\\
2006/05/02	&05:51--06:38	        &13\\
2006/05/22	&02:33--05:07	        &39\\
2006/05/23	&02:11--06:27	        &65\\
2006/06/17	&00:27--01:15	        &13\\
2006/06/18	&00:21--02:10	        &26\\
2006/06/22	&04:02--04:49	        &13\\
2006/07/03	&02:51--03:39	        &13\\
2006/07/17	&00:41--03:13	        &39\\
2006/07/22	&01:24--02:12	        &13\\
2006/07/28	&23:17--23:59	        &13\\
2006/07/29	&00:00--01:54	        &26\\
2006/08/02	&01:12--01:59	        &13\\
2006/08/16	&00:11--01:04	        &13\\
\hline
\end{tabular}
\caption{Log of our observations of \obj\ with the VLT+FORS2 and the 600B grism. All exposure times were 196\,s.}
\label{observations}
\end{center}
\end{table}

Phase-resolved spectroscopy of \obj\ was obtained in `service mode' on a total of 14 nights in May through August of 2006 (see the observation log, table \ref{observations}) with the Very Large Telescope (VLT) of the European Southern Observatory and the FOcal Reducer/low dispersion Spectrograph (FORS2). The observations consist of 312 spectra in the blue--visual range (grism 600B), each having a 196-second exposure time. Median seeing was about $0.8''$, although on a few nights the seeing was significantly worse at $\sim 1.5''$.

All observations were done with a $1''$ slit. The object's celestial position necessitated observations at large zenith angle, where the differential refraction in the atmosphere could not be fully compensated for by the Atmospheric Dispersion Corrector, and the slit was therefore set at parallactic angle for all observations. The detector was the MIT CCD mosaic binned by $2\times2$ pixels, and set to low read-out speed and high gain. The bias subtraction was done by subtracting sixth-order Legendre polynomial fits to the overscan region of each image. A normalised flatfield frame was constructed by averaging the total set of incandescent lamp flatfield frames (5 per night) with rejection of outliers (cosmic rays).

All spectra were extracted using the \textsc{iraf} implementation of optimal (variance-weighted) extraction. Wavelength calibration was done using the standard HeHgCd arc exposures taken during the day. No significant drifts in the arcs were observed over the months of our observations. The residuals of the fourth-order legendre polynomial fit to the arc lines were about 0.1\,\AA. All spectra were transformed to the heliocentric rest-frame prior to analysis.

The spectra were flux-calibrated using the spectrophotometric standard stars LTT\,7379, observed on June 27 and 28, and Feige\,110, observed on June 22, 26, 27 and 28 of 2006. The flux standard spectra were taken with the same $1''$ slit.

\section{Results}

\subsection{Average spectrum}

\begin{figure}
\centering
\includegraphics[angle=270,width=84mm]{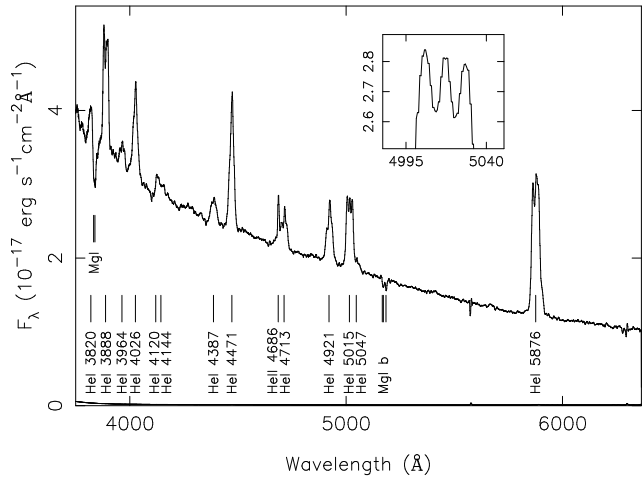}
\caption{Average spectrum of \obj, including line identifications of the helium emission lines and the \mbox{Mg\,{\sc i}} absorption triplets. The lower line, only well visible in the blue part of the spectrum and near skylines due to the Earth's atmosphere in the red part, indicates the estimated noise level. The inset shows a close-up of the triple-peaked profile of the \mbox{He\,{\sc i}} 5015 emission line.}
\label{average}
\end{figure}

The average VLT spectrum of \obj\ is shown in figure \ref{average}. It shows a blue continuum with strong emission lines of helium, very comparable to the long-period AM CVn stars GP Com ($P=46$\,min) and V396 Hya ($P=65$\,min) (e.g.\ \citealt{nather,mtr,lmr}). As in aforementioned systems, the helium lines are triple-peaked, as can best be seen in the \mbox{He\,{\sc i}} 5015 line. The `central spikes' that run in between the common red and blue wings of the emission lines in fact dominate some of the helium emission line profiles, as can be seen in e.g.\ \mbox{He\,{\sc i}} 4471. In addition to the helium emission lines, there is significant absorption from the \mbox{Mg\,{\sc i} {\it b}} triplet around 5175\,\AA. This absorption, which is observed in several of the metal-rich DZ white dwarfs (e.g.\ \citealt{dufour}) and usually attributed to recent or ongoing accretion of metal-rich matter, could be a feature of the accreting white dwarf's photosphere. We subsequently identify the absorption lines at 3833 and 3839\,\AA\ with \mbox{Mg\,{\sc i}} as well (see also \citealt{zuckerman}).

To measure the temperature of the continuum, without the availability of simultaneous and nearby spectrophotometric standard star exposures, we use the following procedure. We create average spectra of \obj\ for three seeing bins: better than $0.6''$, $0.6''-0.7''$, and $0.7''-0.8''$, where we use the seeing as recorded by the seeing monitor at Cerro Paranal. Each of these bins contains about 30 spectra; the spectra taken under worse seeing conditions are not used in order to limit (wavelength-dependent) slit losses as much as possible. For each bin, we assign one (Feige\,110) or two (Feige\,110 \& LTT\,7379) standard star spectra that were taken in comparable seeing conditions, depending on availability. This way the slit losses should -- on average, and as much as possible -- be the same in the binned average spectra and in the standard star spectra. We correct for the differences in airmass using atmospheric extinction data for Cerro Paranal taken from \citet{patat}. This procedure gives a total of 5 flux-calibrated spectra that are more or less independent, and which sample the stochastic variations in for instance the positioning of the stars on the slit as much as possible. We fit a blackbody to these individual flux-calibrated spectra while masking out the obvious spectral lines, and find temperatures between $T_\mathrm{eff}=12,000-15,000$\,K. The hotter temperature fits are obtained for the better seeing bins, where the slit losses should on average be smaller.

In addition to ideal blackbodies we fit helium-atmosphere (DB) white dwarf models from D. Koester, motivated by the fact that the accretor is accreting helium, and the fact that several such systems unambiguously show helium atmospheres (see \citealt{roelofs,roelofsaw}). This yields comparable temperatures, on average about 500\,K lower than pure blackbody fits.

\begin{figure}
\centering
\includegraphics[angle=270,width=84mm]{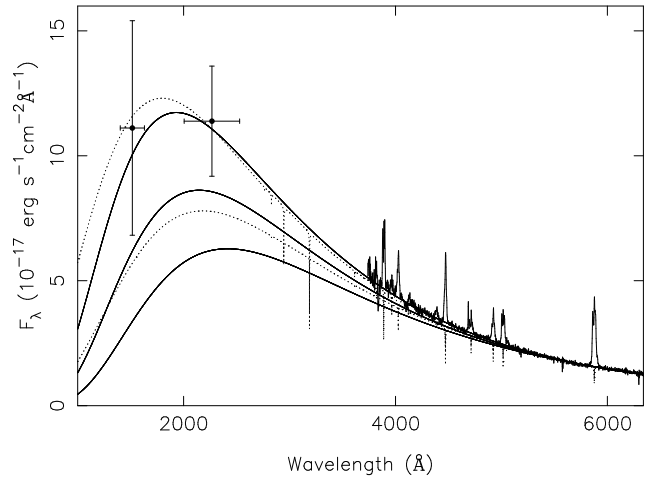}
\caption{Best-seeing flux-calibrated spectrum of \obj\ together with far-UV and near-UV detections from GALEX (solid circles). Top solid line is the best-fitting blackbody temperature of $15,000$\,K, which was fitted to just the optical spectrum. Middle and bottom solid lines indicate $13,500$\,K and $12,000$\,K blackbodies. The upper and lower dotted lines show model atmospheres from D. Koester, with surface gravity $\log g=8$, and $T_\mathrm{eff}=14,000$\,K and $12,000$\,K respectively.}
\label{optical_uv}
\end{figure}

Finally, the broad-band colours from the SDSS (table \ref{systems}) match best with a blackbody of $T_\mathrm{eff}\approx 13,000$\,K (see \citealt{roelofspop}). Broad-band near-UV and far-UV detections of \obj\ in the all-sky survey of the Galaxy Evolution Explorer (GALEX), despite their limited accuracy, favour a $T_\mathrm{eff}\approx 15,000$\,K blackbody or a $T_\mathrm{eff}\approx 14,000$\,K DB white dwarf; see figure \ref{optical_uv}. (These data have not been corrected for Galactic reddening, but the reddening is estimated to be relatively low in the direction of the object, at $E(B-V)=0.025$ mag \citep{schlegel}.) Everything thus points towards an accretor temperature in the range \Teff, if we assume that the optical continuum is dominated by the accretor \citep{bildsten}. The implications of this relatively high temperature are discussed in section \ref{discussion}.

\subsection{The spectroscopic period}

We search for periodicity in the spectra by calculating a time series of flux ratios in the red-shifted and blue-shifted parts of the emission lines, as described in \citet{roelofs}, after \citet{nather}. This has in the past proved to be an effective method for finding the orbital period of such binaries, since the classical `S-wave' signature found in the emission lines of many of these systems causes a modulation of the red-wing/blue-wing flux ratio on the orbital period.

\begin{figure}
\includegraphics[angle=270,width=84mm]{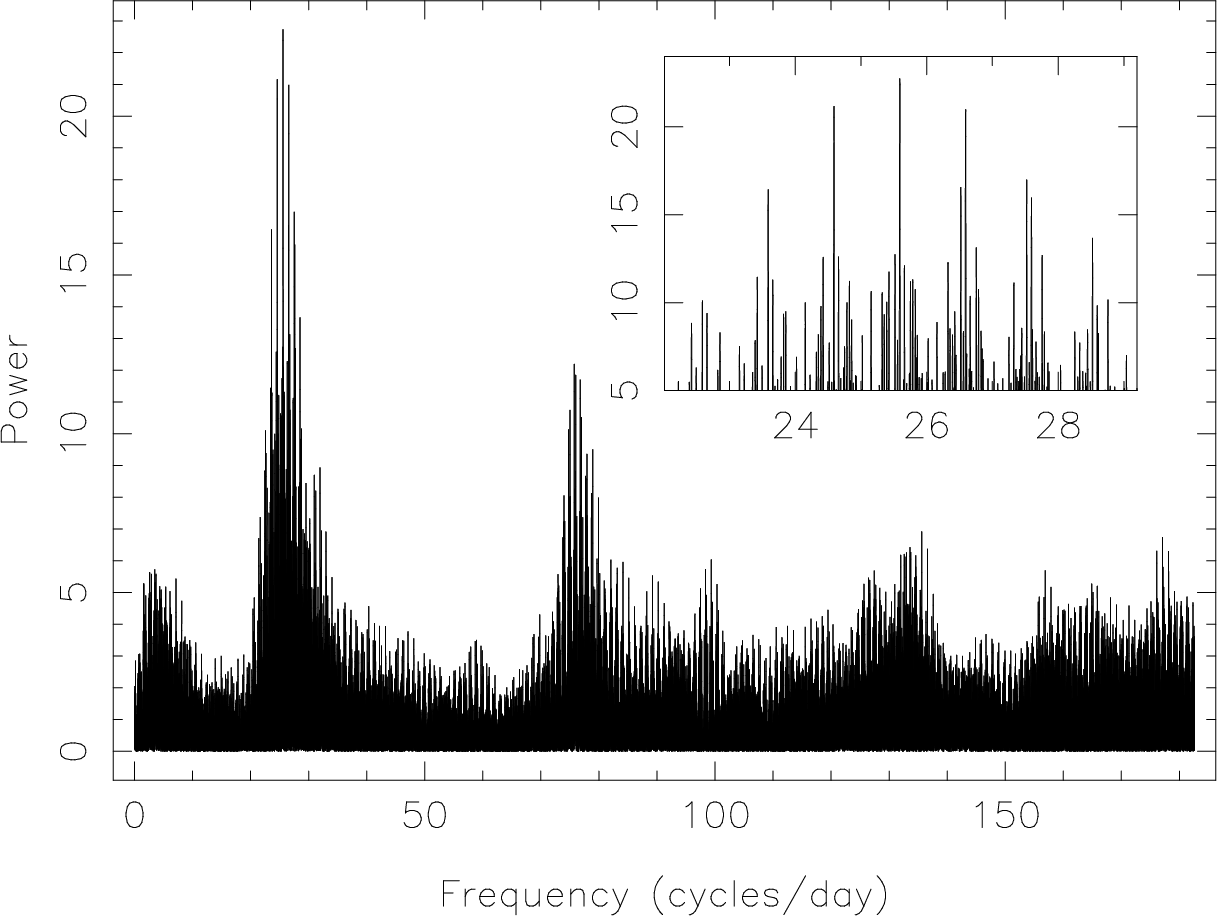}
\caption{Lomb--Scargle periodogram of the red wing/blue wing emission line flux ratios. The inset shows the alias structure around the peak.}
\label{scargle}
\includegraphics[angle=270,width=84mm]{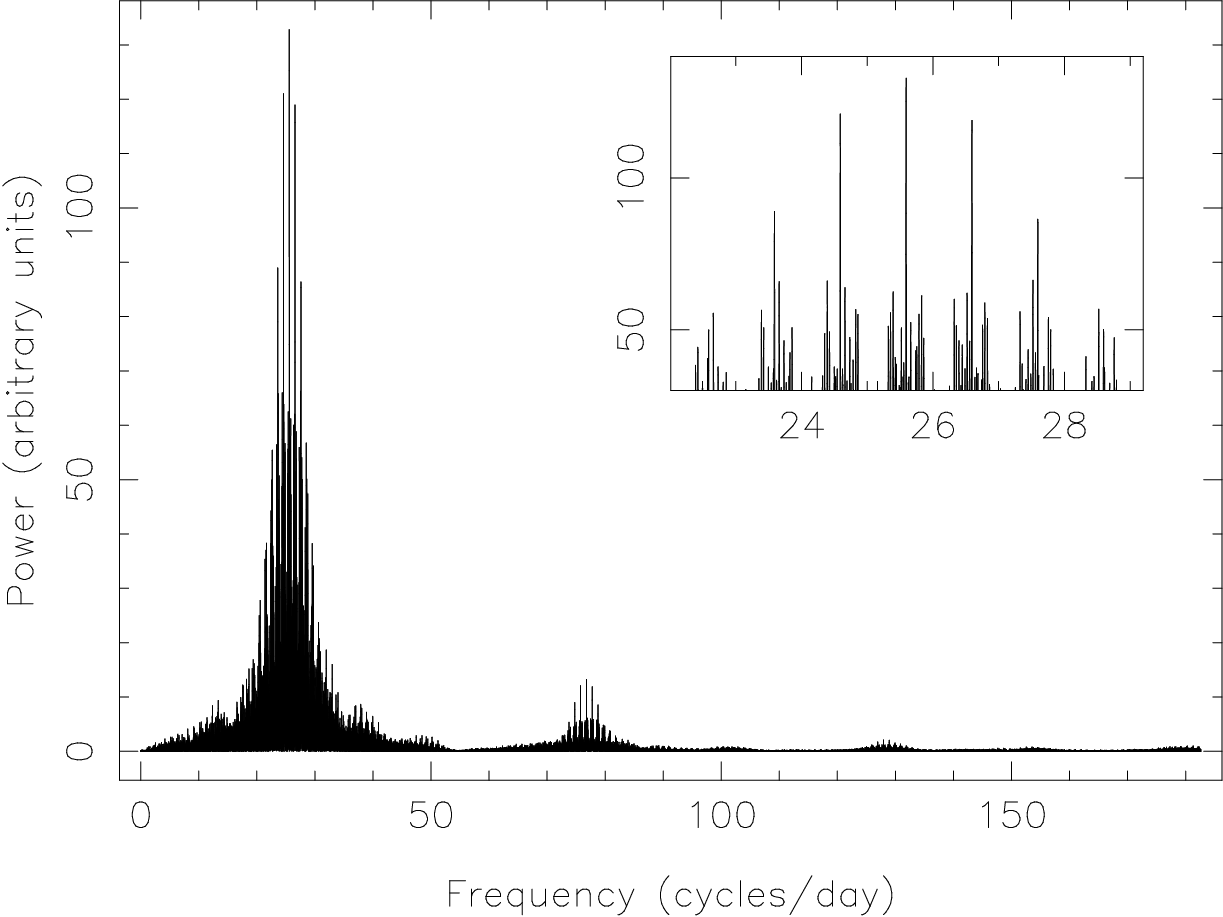}
\caption{Simulated Lomb--Scargle periodogram of the red wing/blue wing emission line flux ratios, for an idealized bright spot (perfect sinusoid, constant visibility) convolved with our actual observations (including phase-smearing due to finite exposure times, but not including noise). This is the `window function'.}
\label{scarglesim}
\end{figure}

A Lomb--Scargle periodogram of this flux ratio versus time is shown in figure \ref{scargle}. It shows a set of highest peaks around 25.59 cycles/day or 3376 seconds, with 1-cycle/day aliases, and a set of third harmonics around 77 cycles/day. Figure \ref{scarglesim} shows the corresponding window function, which was constructed by simulating a bright spot feature in our spectra and calculating, in the same way, a periodogram of red-wing/blue-wing flux ratios. The process of chopping up the emission lines in a red and a blue part discards some of the information on the odd higher harmonics of the test frequency, which causes power to leak into those frequencies as can be seen in the simulated periodogram. The third harmonics, which are commonly observed in such periodograms \citep{roelofs,roelofsaw,roelofsv803}, will thus be (partly) artificial.

\subsection{Dynamic spectrum}

\begin{figure*}
\centering
\includegraphics[angle=270,width=0.9\textwidth]{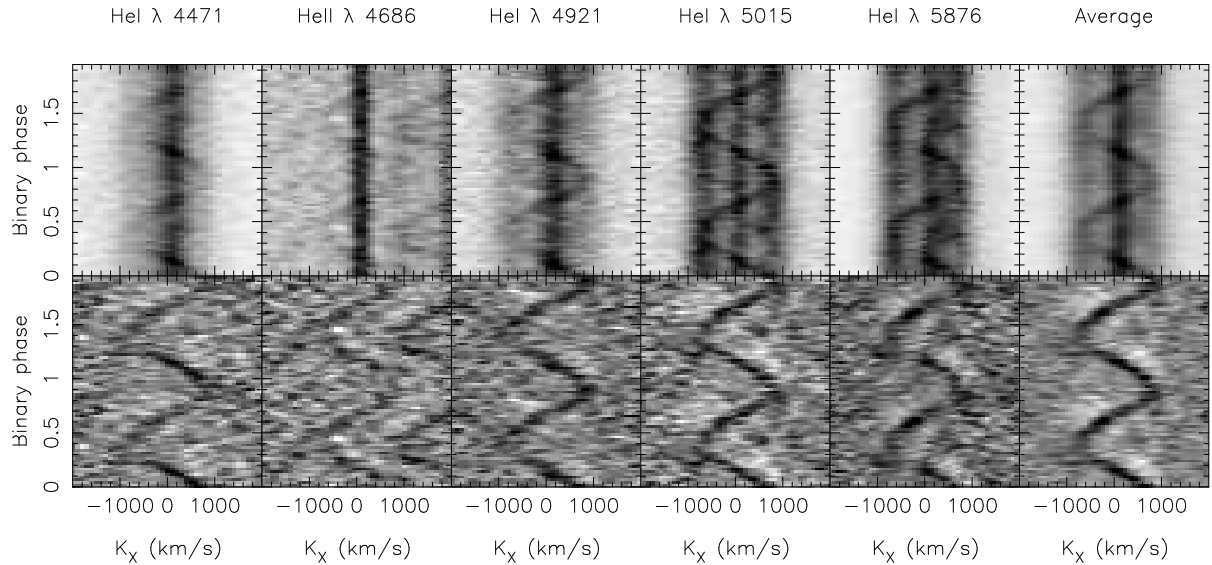}
\caption{Trailed spectra (top row), and average-subtracted trailed spectra of the strongest \mbox{He\,{\sc i}} and \mbox{He\,{\sc ii}} features. The right-most panels show the averages of these spectral lines.}
\label{tomograms}
\end{figure*}

Figure \ref{tomograms} shows the dynamic (trailed) spectra around several spectral lines of helium, after phase-folding the data on the period found from the Lomb--Scargle periodogram in the previous section. Also shown is the average trailed spectrum of these spectral lines, which provides a better signal-to-noise ratio. A classical double-peaked accretion disk line profile plus a narrow emission component near zero velocity is observed, quite like the other emission-line AM CVn stars (e.g., \citealt{nather,mtr,lmr,roelofs}). In most lines, and most obviously in the combination of several helium lines, the `S-wave' component is clearly visible. This component is usually attributed to an enhanced emission at the site where the accretion stream impacts the disk, and supports our identification of the correct orbital period of the binary (for the stream--disc impact point should rotate with the binary).

From the S-wave signal and the requirement of its coherence in a phase-folded trailed spectrum, we derive an orbital period of \Porb. When phase-folded on a period that is off by about a quarter of an S-wave cycle over the four-month baseline of our observations, the S-wave signal starts to lose coherence and disappears from the trailed spectrum, which gives us a direct estimate of the error on the orbital period. Neighbouring 1/day aliases, which show up relatively strongly in the Lomb--Scargle periodogram of figure \ref{scargle}, are effectively ruled out by phase-folding the spectra on these test frequencies, and verifying that the S-wave signal is significantly less pronounced at and around these frequencies.

\section{Discussion}
\label{discussion}

We have presented an orbital period and effective temperature measurement of the recently discovered AM CVn star candidate \obj\ \citep{anderson}. The orbital period of \Porb\ ($56.272\pm0.005$\,min) confirms that the source is an ultracompact binary star, and our spectra confirm its hydrogen-deficient nature. We can thus classify \obj\ as a new member of the AM CVn class.

Of particular interest for long-period AM CVn stars is the question of the presence of hydrogen, since one of the proposed formation channels for AM CVn stars (the `evolved-Cataclysmic-Variable channel'; \citealt{podsi}) predicts that a disproportionately large fraction of AM CVn-like binaries from this channel reside at orbital periods not too far below the orbital period minimum for hydrogen-rich donors (which lies at about 80 minutes), the majority of which should contain some hydrogen. A few hydrogen-rich Cataclysmic Variables (CVs) with slightly evolved donors are known, the shortest-period one being V485 Cen at $P_\mathrm{orb}=59$\,min \citep{tau}. But in strong contrast with V485 Cen, which has a hydrogen-dominated spectrum, there is no evidence for hydrogen emission or absorption in the average or in the dynamic spectrum of \obj. Based on a simple LTE model for a slab of helium-rich gas (described in \citealt{trm91}), assuming a temperature of around $\sim$11,000\,K in the region where the (helium) emission lines are formed \citep{trm91}, we tentatively constrain the hydrogen content to $\mathrm{H/He}\lesssim 10^{-4}$ by mass. From the apparent absence of hydrogen in the spectrum we can say at least that \obj\ is not an evolved CV currently moving towards shorter orbital periods, since its hydrogen content would have to be at least similar to that of V485 Cen. It could, however, still be a binary from the evolved-CV channel if it is currently evolving away from a much shorter orbital period after fully depleting hydrogen in the donor star's core. But since the number of possible evolutionary solutions in which there should still be an easily detectable fraction of hydrogen (say, at least 1\%) increases with orbital period, this becomes less likely for longer-period systems.

For the continuum in \obj's spectrum, we have measured a temperature between \Teff. Since the optical continuum is expected to be dominated by the accreting white dwarf in such long-period AM CVn stars \citep{bildsten}, the measured temperature most likely reflects the temperature of the accreting white dwarf. It is then very interesting to note that this temperature is at the upper end of the range predicted by recent models for the thermal evolution of the accreting white dwarfs in AM CVn binaries (see figure 2 in \citealt{bildsten}). Essentially, more massive components at a given orbital period imply higher mass transfer rates as set by gravitational-wave radiation, and the resulting stronger accretion heating gives a hotter accretor. Our temperature for \obj\ is best compatible with the evolutionary cooling track of a hot donor star paired with a relatively massive accretor of $M_1\gtrsim 1.0\,M_\odot$ (see \citealt{bildsten}).

More detailed models for the thermal evolution of the donor stars in AM CVns, which allow the donors to cool radiatively \citep{donorevolution}, indicate that AM CVns with hot donors evolve to orbital periods of $\sim$1\,h a bit more slowly than the adiabatic models of \citet{deloye} used in \citet{bildsten}, giving the accretors more time to cool. This would push our measured temperature even more toward the high-mass end of the predicted range, as compared to \citet{bildsten}.

Although it is rather indirect evidence, it is an interesting result in the light of recent findings that the donor stars in many short-period AM CVn stars appear to be hotter and more massive than they should be if they were fully degenerate (\citealt{roelofsamcvn,roelofshst,marsh0926}; see also \citealt{nasser}). The accretor's temperature at long orbital periods may provide a completely alternative probe of the binary's component masses when it was still at much shorter orbital periods, under the crucial assumption that the temperature observed today represents the accretor's longer-term equilibrium temperature as predicted by the cooling models. This assumption could break down when phases of enhanced accretion temporarily raise the accretor's effective temperature (e.g.\ \citealt{piro}).

We furthermore cannot exclude that the accretion luminosity contributes to the continuum of the observed spectrum, even though the present-day accretion luminosity (as set by gravitational-wave radiation) is expected to be small compared to the luminosity of the accretor, at these long orbital periods (e.g.\ \citealt{bildsten}). Even for an $M_1=1.0 M_\odot$ accretor, gravitational-wave radiation is only able to power an accretion-disc-sized blackbody of $T_\mathrm{eff}\sim 3500$\,K, depending on the mass of the donor star and on the exact size of the disc (e.g.\ \citealt{roelofshst}). It thus appears unlikely that a contribution from the disc could noticeably increase the observed continuum temperature in such a system, even though the disc's true spectral energy distribution will deviate from a single-temperature blackbody.

A question to be discussed is whether our measured temperature may be erroneously high; after all, our spectrophotometric standard star observations were less than ideal, and wavelength-dependent slit losses may have affected the flux calibrations. However, since \obj\ was observed at higher airmass than the standard stars, while we have assigned standard star spectra based on near-zenith seeing, it can be expected that the blue flux in our spectra has been suppressed due to worse seeing at higher airmass. This would cause our measurement of $T_\mathrm{eff}$ to be on the low rather than on the high side (and more strongly so for the worse-seeing spectra, which gave the lower temperatures). We thus consider it unlikely that we are overestimating the continuum temperature. Lastly, we note that if one assumes an accretor temperature much \emph{higher} than $15,000$\,K, one would expect to see the helium absorption lines of the accretor's photosphere, as clearly observed in the shorter-period AM CVn stars SDSS\,J1240$-$0159 and V406 Hya \citep{roelofs,roelofsaw}.

\section{Acknowledgments}

GHAR is supported by NWO Rubicon grant 680.50.0610 to G.H.A. Roelofs. PJG is supported by NWO VIDI grant 639.042.201 to P.J. Groot. GN is supported by NWO VENI grant 639.041.405 to G. Nelemans. DS acknowledges support through the NASA Guest Observer program and is supported by a PPARC/STFC Advanced Fellowship. This work is based on data taken at the European Southern Observatory, Chile, under programme 077.D-0686(A). This research was supported in part by the National Science Foundation under Grant No. PHY05-51164. The authors acknowledge the hospitality of the Kavli Institute for Theoretical Physics at the University of California, Santa Barbara. D. Koester is acknowledged for kindly making available his white dwarf atmosphere models.

\end{document}